\documentclass[conference]{IEEEtran}
\usepackage{graphicx}

\ifCLASSINFOpdf
\else
\fi
\begin{document}
\IEEEoverridecommandlockouts
%
\title{AutoPhase: Compiler Phase-Ordering for HLS 
with Deep Reinforcement Learning}

 \author{Ameer Haj-Ali*, Qijing Huang*, William Moses, John Xiang, Ion Stoica, Krste Asanovic, \\ John Wawrzynek \\

 \{ameerh, qijing.huang, johnxiang, istoica, krste, johnw\}@berkeley.edu, wmoses@mit.edu \\
\thanks{\hspace{-0.27cm}\rule{3.15in}{0.4pt}}
\thanks{\fontsize{9.5}{13.5}\selectfont \textbf{* Equal contribution.}}
 }


%


\maketitle
\begin{abstract}
The performance of the code generated by a compiler depends on the order in which the optimization passes are applied.  In high-level synthesis, the quality of the generated circuit relates directly to the code generated by the front-end compiler.
Choosing a good order--often referred to as the {\em phase-ordering} problem--is an NP-hard problem. 
In this paper, we evaluate a new technique to address the phase-ordering problem: deep reinforcement learning.
We implement a framework in the context of the LLVM compiler to optimize the ordering for HLS programs and compare the performance of deep reinforcement learning to state-of-the-art algorithms that address the phase-ordering problem.
Overall, our framework runs one to two orders of magnitude faster than these algorithms, and achieves a 16\% improvement in circuit performance over the -O3 compiler flag.
\end{abstract}

%
\IEEEpeerreviewmaketitle
\section{Introduction} 
\label{sec:bg}
Compilers execute optimizations to transform programs into more efficient forms to run on various hardware targets.
Different phase orderings can generate different low level codes that significantly differ in performance. Multiple machine learning approaches have been proposed to overcome this challenge. Recent  advancements in deep reinforcement learning (RL) ~\cite{kaelbling1996reinforcement}
offer opportunities to address the phase ordering challenge. In RL, a learning agent observes the state of the environment, and takes an action based on this observation. The goal of the RL agent is to compute a policy--a mapping between the environment states and actions--that maximizes a long-term reward.  In this work, we focused on two RL algorithms: Policy Gradient (PG) and Deep Q-Network  (DQN).
\section{Our Framework and Results} 
\label{sec:int}
We developed analysis passes to extract 56 static features from the LLVM program intermediate representation, such as the number of basic blocks, branches, and instructions of various types. 
We represent the states in RL either with these features or a histogram of applied passes. Both representations are effective. 
We use the number of clock cycles reported by the HLS profiler from LegUp \cite{canis2011legup} as the performance metric. Success is defined as learning a sequence of passes that performs better than -O3 within a reasonable time. We use $12$ HLS benchmarks for evaluation taken from CHstone~\cite{chstone_short} and LegUp examples.   
We ran DQN and PG, and compared it against random search, greedy algorithms, genetic algorithms, and -O3. 
\begin{figure}[!t]
    \centering
    \includegraphics[width=8.5cm]{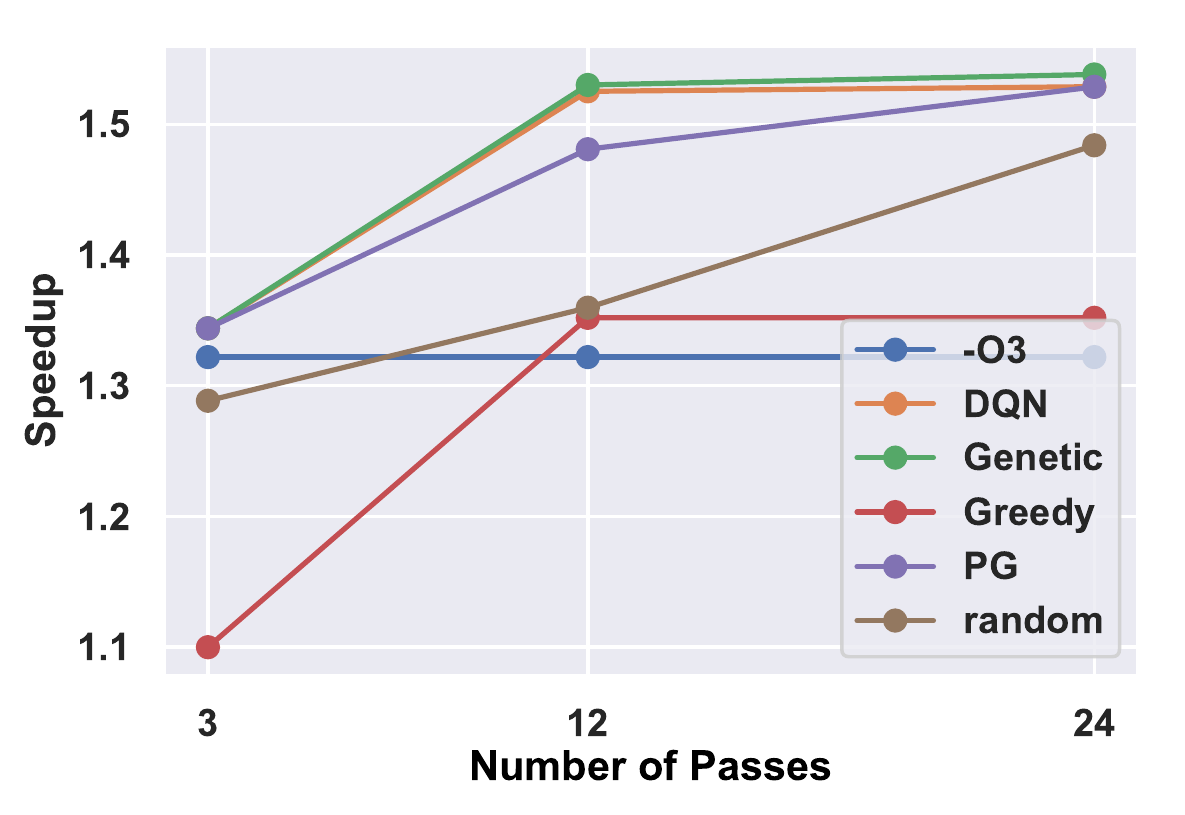}
    \caption{Circuit speedup of various algorithms compared to No-Opt with sequence length of 3, 12, and 24.}
    \label{fig:length}
\end{figure}
Figure \ref{fig:length} shows the circuit speedup for various algorithms with sequence lengths of 3, 12, and 24.  
Little improvement is observed by increasing the number of applied passes from 12 to 24. Overall, RL and Genetic algorithms achieve the highest circuit speedups -- 16\% better than -O3. While RL and Genetic algorithms achieved similar performance, RL runs $3\times$ faster. Furthermore, a trained RL agent could be used to find optimal passes for other programs with little retraining and further improves the algorithm runtime.

\section{Conclusion} 
\label{sec:conc}
In this work, we explore a novel deep RL based approach to improve performance of HLS designs by optimizing the compiler phase ordering. 
Our preliminary results show that RL techniques achieve 16\% better results than traditional -O3. Such improvement requires only a few minutes of training. Future work includes improving generalization in RL. 


%

\bibliographystyle{IEEEtran}
\bibliography{projBib}

\begin{thebibliography}{1}
\providecommand{\url}[1]{#1}
\csname url@samestyle\endcsname
\providecommand{\newblock}{\relax}
\providecommand{\bibinfo}[2]{#2}
\providecommand{\BIBentrySTDinterwordspacing}{\spaceskip=0pt\relax}
\providecommand{\BIBentryALTinterwordstretchfactor}{4}
\providecommand{\BIBentryALTinterwordspacing}{\spaceskip=\fontdimen2\font plus
\BIBentryALTinterwordstretchfactor\fontdimen3\font minus
  \fontdimen4\font\relax}
\providecommand{\BIBforeignlanguage}[2]{{%
\expandafter\ifx\csname l@#1\endcsname\relax
\typeout{** WARNING: IEEEtran.bst: No hyphenation pattern has been}%
\typeout{** loaded for the language `#1'. Using the pattern for}%
\typeout{** the default language instead.}%
\else
\language=\csname l@#1\endcsname
\fi
#2}}
\providecommand{\BIBdecl}{\relax}
\BIBdecl

\bibitem{kaelbling1996reinforcement}
L.~K. \textit{et al.}, ``Reinforcement learning: A survey,'' vol.~4, 1996, pp.
  237--285.

\bibitem{canis2011legup}
A.~C. \textit{et al.}, ``{LegUp: high-level synthesis for FPGA-based
  processor/accelerator systems},'' in \emph{FPGA 2011}, pp. 33--36.

\bibitem{chstone_short}
Y.~H. \textit{et al.}, ``{CHstone: A benchmark program suite for practical
  c-based high-level synthesis},'' in \emph{ISCAS 2008}, pp. 1192--1195.

\end{thebibliography}

\end{document}